# Fully Polarimetric SAR and Single-Polarization SAR Image Fusion Network

Liupeng Lin[a], Jie Li[b], Huanfeng Shen[a,c], Lingli Zhao[d], Qiangqiang Yuan[b,c], Xinghua Li[d]

[a] *School of Resource and Environmental Sciences, Wuhan University, Wuhan, China.*

[b] *School of Geodesy and Geomatics, Wuhan University, Wuhan, China.*

[c] *Collaborative Innovation Center of Geospatial Technology, Wuhan University, Wuhan, China.*

[d] *School of Remote Sensing and Information Engineering, Wuhan University, Wuhan, China.*

[*] *Corresponding author, E-mail address: shenhf@whu.edu.cn*

**Abstract**

The data fusion technology aims to aggregate the characteristics of different data and obtain products with multiple data advantages. To solves the problem of reduced resolution of PolSAR images due to system limitations, we propose a fully polarimetric synthetic aperture radar (PolSAR) images and single-polarization synthetic aperture radar SAR (SinSAR) images fusion network to generate high-resolution PolSAR (HR-PolSAR) images. To take advantage of the polarimetric information of the low-resolution PolSAR (LR-PolSAR) image and the spatial information of the high-resolution single-polarization SAR (HR-SinSAR) image, we propose a fusion framework for joint LR-PolSAR image and HR-SinSAR image and design a cross-attention mechanism to extract features from the joint input data. Besides, based on the physical imaging mechanism, we designed the PolSAR polarimetric loss function for constrained network training. The experimental results confirm the superiority of fusion network over traditional algorithms. The average PSNR is increased by more than 3.6db, and the average MAE is reduced to less than 0.07. Experiments on polarimetric decomposition and polarimetric signature show that it maintains polarimetric information well.

## 1. INTRODUCTION

The PolSAR image imaging system can obtain rich texture information and polarimetric information under day-and-night and all-weather conditions, so it has been widely used in the field of geosciences, such as land use and land cover[1]–[3], ship detection[4], [5], and buildings damage assessment[6], [7], crop monitoring[8], [9], wetland classification[10], etc. However, due to the complexity of the SAR imaging system and the limitations of hardware equipment, it is inevitable to the trade-off between polarimetric information, image swath, and image resolution, when acquiring SAR images. For some SAR sensors, to obtain PolSAR images with multiple polarimetric information, the resolution of the image is reduced. This case means that the resolution of PolSAR images acquired by these sensors is often lower than that of SinSAR images. The reduction in resolution will further affect its application in practice. To solve the problem of low resolution of PolSAR images, some scholars use frequency-domain methods to enhance spatial resolution. In 2001, Debora Pastina et al.[11] first proposed to introduce polarimetric information into the super-resolution of PolSAR images. They used SPECAN techniques to perform single-channel super-resolution of PolSAR images to improve the spatial resolution. Suwa et al.[12] proposed a 2D-PBWE method by extending the traditional bandwidth extrapolation method from SAR images to PolSAR images. By using the 2D linear model, they extrapolated spatial frequency bandwidth from the azimuth direction and the distance direction to improve spatial resolution. Those methods may be able to reconstruct a higher resolution, but the utilization of polarimetric information is insufficient.

Another part of scholars uses the prior information or characteristics of the image itself to enhance the PolSAR image resolution. Yang et al.[13] used the POCS algorithm to construct a convex set to extract information from different polarimetric channels of low-resolution SAR images. The high-resolution polarimetric image for ship detection is generated by fusing different polarimetric channel information, but this method discards the polarimetric and phase information of original SAR image during the processing[14]. Zou et al.[15] proposed SRQP to improve PolSAR image spatial resolution. First, PolSAR image is decomposed into multiple polarimetric components, and then uses the adjacent pixels of the center pixel of each polarization component to perform a weighted evaluation on the sub-pixels of the center pixel. And through polarimetric synthesis to obtain polarimetric images. This method only considers the spatial correlation between a pixel and its neighbors, resulting in an obvious grid effect. To solve the grid effect problems of SRQP, the SRPSC[14] method was proposed. This method uses the polarimetric spatial correlation between pixels to determine the initial value of the sub-pixel, and iteratively solves it to obtain a convergent HR-PolSAR image. Based on polarimetric coherent target decomposition, this method can maintain the polarimetric information well. However, the super-resolution accuracy will be limited by inappropriate polarimetric decomposition.

In recent years, deep learning is widely used in image reconstruction and restoration tasks[16]–[22]. To date, there are relatively few PolSAR super-resolution algorithms using deep learning. For the first time, MSSR method[23] is proposed to introduce deep learning into the PolSAR images super-resolution reconstruction, which is used for processing multi-channel PolSAR data simultaneously. However, it does not fully consider the polarimetric information and numerical characteristics of PolSAR images. Subsequently, PSSR method[24] is proposed to solve this problem. They used a complex block, transposed convolution, PReLU, and other structures to maintain polarimetric and numerical characteristics. This method can effectively enhance PolSAR image resolution and maintain polarimetric information.

In this paper, inspired by fusion technology[25]–[27], we propose a novel PolSAR image and SinSAR image fusion network (PSFN) to efficiently obtain HR-PolSAR images. We adopted a novel data fusion framework and designed the corresponding network structure according to the characteristics of the data. Besides, we designed a polarimetric loss function based on the physical imaging mechanism of the PolSAR data. The contributions of PSFN are:

1) Fusion framework for LR-PolSAR images and HR-SinSAR images. Different from the existing PolSAR image resolution reconstruction method, this strategy can fully integrate spatial texture information of the SinSAR image while maintaining polarimetric information of the PolSAR image.

2) Novel PolSAR attention mechanism. We designed the cross-attention mechanism to take advantage of polarimetric information of LR-PolSAR images and spatial information of HR-SinSAR images.

3) Loss functions based on the PolSAR physical imaging mechanism. The polarimetric loss function is used to intuitively constrain network training and enhance network interpretability.

The rest of this paper is organized as follows. Section 2 introduces the PolSAR data and the detailed architecture of PSFN. In Sections 3, The experimental results on four datasets are presented. The polarimetric analysis is provided in Section 4. The conclusions and prospects for future research are summarized in Section 5.

## 2. DETAILS OF THE PROPOSED FUSION NETWORK

### 2.1. PolSAR Data Organization

Through multiple transmission and reception modes of echo, the PolSAR imaging system can obtain a backscattering matrix composed of multiple polarimetric channels with different polarimetric characteristics. For monostatic PolSAR data, it can be expressed in the form of a $2 \times 2$ complex backscattering matrix[28], as shown in the following equation:

$$S = \begin{bmatrix} S_{HH} & S_{HV} \\ S_{VH} & S_{VV} \end{bmatrix} \tag{1}$$

where $S_{HH}$ and $S_{VV}$ represent the co-polarized channels, and $S_{HV}$, $S_{VH}$ represent the cross-polarized channels. SinSAR data refers to SAR data that contains only one polarimetric mode among $S_{HH}$, $S_{VV}$, $S_{HV}$, and $S_{VH}$, while fully PolSAR data refers to SAR data that contains all four polarimetric modes.

In the case of satisfying the reciprocity theory and ignoring the influence of system noise, the back-scattering matrix can be vectorized into the Lexicographic covariance matrix $C_3$ [28], [29].

$$C_3 = \begin{bmatrix} |S_{HH}|^2 & \sqrt{2}S_{HH}S_{HV}^* & S_{HH}S_{VV}^* \\ \sqrt{2}S_{HV}S_{HH}^* & 2|S_{HV}|^2 & \sqrt{2}S_{HV}S_{VV}^* \\ S_{VV}S_{HH}^* & \sqrt{2}S_{VV}S_{HV}^* & |S_{VV}|^2 \end{bmatrix} \quad (2)$$

where, $*$ represents the conjugate operation. Since the two elements in the off-diagonal, symmetric position of the $C_3$ is conjugate complex numbers, it can be converted into a $1 \times 9$ real-valued vector.

$$C_{value} = [R_{11}, R_{12}, I_{12}, R_{13}, I_{13}, R_{22}, R_{23}, I_{23}, R_{33}]^T \quad (3)$$

Among them, $R$ represents the value of real part, $I$ represents the value of imaginary part, and the subscript represents its position in the $C_3$ matrix. For a scene of HR-PolSAR image $C_x$, due to system limitations and other factors, it degenerates into the observational LR-PolSAR image $C_y$. The degradation model is $C_y = f_d(C_x)$. $f_d(.)$ is down-sampling operator, which used to describe degradation process between HR-PolSAR images and LR-PolSAR images. Among them, both HR-PolSAR image $C_x$ and LR-PolSAR image $C_y$ are three-dimensional matrices composed of $C_{value}$. $C_{value}$ represents the value of one pixel of $C_x$ and $C_y$. Besides, the relationship between HR-SinSAR intensity image and HR-PolSAR image can be expressed as: $I_{x,i} = \Im_i(C_x), i \in (HH, HV, VH, VV)$. $\Im(.)$ represents the intensity image extraction operator.

## 2.2. Fusion framework for PolSAR and SinSAR

To overcome the reduction of PolSAR image resolution caused by the degradation process, we propose a novel PolSAR image and SinSAR image fusion network. As illustrated in Fig. 1, The proposed fusion network is the end-to-end residual convolutional neural network. The input of the proposed method is the joint input of LR-PolSAR data and HR-SinSAR auxiliary data. The spatial features of HR-SinSAR image and polarimetric features of LR-PolSAR image are extracted by the HR-SinSAR feature extraction module (HSFE) and LR-PolSAR super-resolution module (LPSR) respectively. Besides, a cross-attention mechanism (CroAM) is designed to weigh the input data mutually and guide extraction of spatial information and polarization information, which includes two sub-modules: HR-SinSAR spatial attention module (HSSA) and LR-PolSAR channel attention module (LPCA). Among them, the HSEF module and LPSR module separately extract information from HR-SinSAR and LR-PolSAR, while the CroAM module uses HR-SinSAR and LR-PolSAR to extract information under mutual guidance. Through the separate and cooperative information extraction modules, while ensuring the effective information extraction of a single data, the guided spatial information and polarization information is extracted. In terms of the loss function, according to the physical imaging mechanism, a corresponding polarimetric loss function is designed to constrain the training process.

For this fusion framework, we use LR-PolSAR image $C_y \in \mathbb{R}^{hei \times wid \times cha}$ and HR-SinSAR intensity image $I_x \in \mathbb{R}^{HEI \times WID \times 1}$ as the input to the network, and generate HR-PolSAR image $C_x \in \mathbb{R}^{HEI \times WID \times CHA}$ through the proposed network. Among them, $HEI$, $WID$, $CHA$ is height, width, and channel of HR-PolSAR image, respectively. $r_{azimuth} = HEI / hei$ and $r_{range} = WID / wid$ are the spatial resolution ratios of LR-PolSAR image and HR-PolSAR image in the azimuth and distance direction, respectively. HR-SinSAR image and LR-PolSAR image are used as the only input for feature extraction of the HSFE module and LPSR module, respectively. In the HSSA module and the LPCA module, the HR-SinSAR image and the LR-PolSAR image are simultaneously used as joint input data. In the fusion framework, the accurate polarimetric information of the PolSAR image is used to correct the feature maps of the SinSAR, and the high-precision spatial details of the SinSAR are used to enhance spatial information of PolSAR image. The implementation details is provided below.

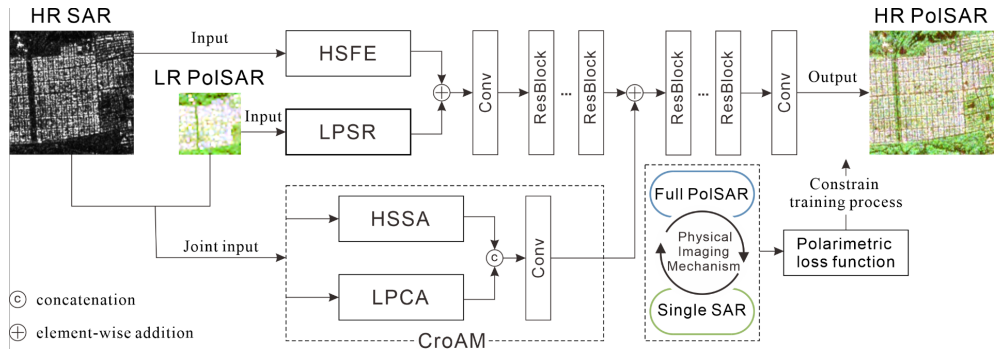

**Fig. 1.** The flowchart of the proposed method.

### 2.3. High-resolution Single-polarization SAR Feature Extraction Module

As displayed in Fig. 2, in HR-SinSAR image feature extraction module, firstly, one convolution layer is used to increase the dimension of the SinSAR intensity image to fully extract spatial information of SinSAR image. In this module, five residual units are used to extract features. The residual structure[30] was proposed to solve the problem of performance degradation in the deep network and accelerate network convergence[31]. In this paper, the residual unit consists of three parts: two-dimensional convolution, PReLU activation function, and residual structure. The residual unit uses two-dimensional convolutional layers to extract features, uses the PReLU activation function to enhance nonlinear characteristics, and uses the residual structure to connect the convolutional layers. The residual unit structure is illustrated in Fig. 3. The attention mechanism is widely used in computer vision tasks to recalibrate feature maps[32]–[38]. The spatial attention mechanism can recalibrate the high-frequency parts of the image with rich texture information, thereby enhancing the ability to extract spatial detail information. In the proposed method, a spatial attention module is introduced into the HSFE module, which aims to better extract the spatial detail information of the SinSAR image. As shown in Fig. 4, in the spatial attention module, one convolution layer is used to distill the information in the channel direction of the feature maps extracted by the residual unit, and convert it into a two-dimensional feature map $\in \mathbb{R}^{H \times W \times 1}$. Subsequently, the sigmoid function is adopted to normalize feature maps to obtain spatial weight layer. Finally, the spatial weight layer is multiplied by the residual unit feature maps as the weighted feature maps of the spatial attention mechanism. The spatial attention module can be expressed as:

$$F_{SA} = F_{input} \odot \sigma\left(W_{SA} \circ F_{input} + b_{SA}\right) \qquad (4)$$

where, $F_{SA}$ is output feature maps, $F_{input}$ is input feature maps. $W_{SA}$ is the convolution kernel of the spatial attention module, $b_{SA}$ represents the bias of it, $\circ$ is convolution operation, and $\sigma(.)$ represents the sigmoid activation function, $\odot$ is element-wise multiplication function. For the feature maps extracted by the HSFE module, we directly add them as auxiliary information to the LPSR module to enhance the spatial detail information of the LR-PolSAR super-resolution module.

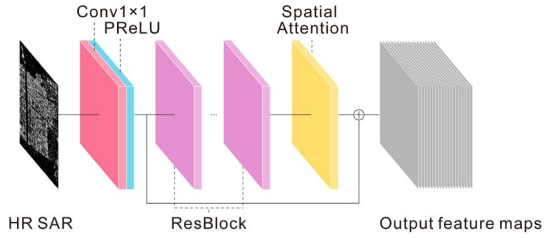

**Fig. 2.** The HSFE module.

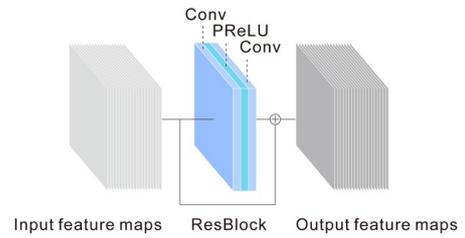

**Fig. 3.** Residual unit.

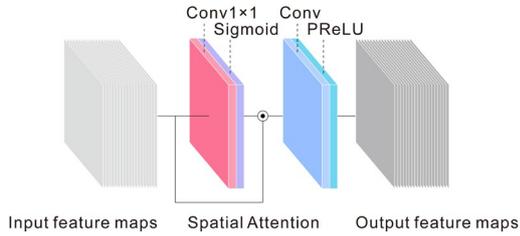

**Fig. 4.** The spatial attention module.

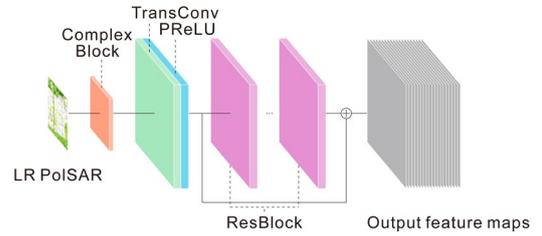

**Fig. 5.** The LPSR module.

### 2.4. Low-resolution PolSAR Super Resolution Module

In the LR-PolSAR image super-resolution reconstruction module, we make a series of improvements to the PSSR network. As shown in Fig. 5, deconvolution is used to up-sample the LR-PolSAR image to improve its resolution and keep the complex blocks[24] in the PSSR network which are used to extract the PolSAR image mixed characteristic. Also, several residual units with the same configuration as in the HSFE module are used to replace the 20 convolutional layers in the PSSR network to speed up the training process of the super-resolution module.

In the PSSR network, deconvolution [39] is employed to up-sample the LR-PolSAR image first, and then complex blocks are used to extract mixed features from the up-sampled PolSAR image. In the case of insufficient convergence in the initial training stage, the process that first uses transposed convolution to up-sample the image and then employs the complex block to extract features may cause a certain error in the mixed feature extracted from the complex block. Therefore, in the proposed method, we use complex blocks to extract mixed features from LR-PolSAR images and then use transposed convolution to up-sample the extracted feature maps. In this way, the accuracy of the feature maps obtained at initial training stage is improved.

It should be noted that, different from HFSE module, this fusion network does not use any attention module in the LPSR module. The essence of attention module is to assign a higher weight to the information of interest in the image, which will lead to the phenomenon that the information in some areas of the image is weakened. Our goal is to enhance the spatial texture information of LR-PolSAR images while maintaining its original polarimetric information. Therefore, we did not introduce the attention mechanism in the LPSR module, but used the output of the LPSR module as the basic information.

## 2.5. Cross-Attention Mechanism

To take advantage of the complementary information of LR-PolSAR image and HR-SinSAR image, a cross-attention mechanism is designed to use respective advantages of LR-PolSAR data and HR-SinSAR data to enhance the polarimetric information and spatial texture information of feature maps generated by the LPSR module. The mechanism consists of two parts: a high-resolution single-polarization SAR spatial attention module and a low-resolution fully polarimetric SAR channel attention module.

### 2.5.1 High-resolution Single-polarization SAR Spatial Attention Module

Compared with LR-PolSAR images, HR-SinSAR images have higher accuracy in spatial information. Using its high-precision spatial information to guide and weight the feature maps extracted from LR-PolSAR images can make the spatial features extracted by LR-PolSAR more accurate and reliable. The HSSA module aims to extract the spatial information weights of HR-SinSAR images and use the weights to recalibrate the feature maps of LR-PolSAR. The spatial information of HR-SinSAR is introduced to enhance LR-PolSAR image. The structure diagram of the HSSA module is displayed in Fig. 6.

In the HSSA module, one convolutional layer is employed to reduce the dimensions of feature maps of HR-SinSAR images along the channel direction and then normalize them through the sigmoid function to generate a spatial weight layer. On the other hand, we use transposed convolution to up-sample the LR-PolSAR image to $\mathbb{R}^{H \times W \times S}$. Subsequently, we multiply the spatial weight layer with the feature maps of the up-sampled PolSAR image to obtain the feature maps of the spatially weighted PolSAR image. The output of HSSA can be written as follows:

$$F_{HSSA} = (W_{HSSA1} \circ F_{LP} + b_{HSSA1}) \odot \sigma(W_{HSSA2} \circ F_{HS} + b_{HSSA2}) \tag{5}$$

where, $F_{HSSA}$ is output feature maps of HSSA module, $F_{LP}$ and $F_{HS}$ respectively represent the feature maps generated by LR-PolSAR and HR-SinSAR. $W_{HSSA1}$ and $W_{HSSA2}$ represent the convolution kernel of the HSSA module, $b_{HSSA1}$ and $b_{HSSA2}$ represent the bias of the HSSA module.

### 2.5.2 LR-PolSAR Channel Attention Module

Compared with HR-SinSAR images, LR-PolSAR images have richer polarization information. If the PolSAR image is directly generated from a SinSAR image, it is obvious that the generated PolSAR image has lower reliability in terms of polarimetric information. Therefore, we use the accurate polarimetric information of LR-PolSAR images to guide the generation of feature maps of SinSAR images, so that the feature maps extracted by HR-SinSAR have both high-precision polarimetric information and spatial texture information. The LPCA module is designed to extract the polarimetric information weights of LR-PolSAR images and guide the feature maps of HR-SinSAR. The polarimetric information of PolSAR is introduced to enhance the HR-SinSAR image.

As shown in Fig. 7, in the LPCA module, we use transposed convolution to up-sample the LR-PolSAR image and extract feature maps, and then use the channel attention mechanism to generate the channel weight layer. Subsequently, it is multiplied by the feature maps extracted from HR-SinSAR images to obtain feature maps of channel-weighted SinSAR images. Through the channel attention module, LR-PolSAR images are used to enhance the polarimetric information of SinSAR images.

The LPCA module can be written as follows:

$$F_{LPCA} = (W_{LPCA1} \circ F_{HS} + b_{LPCA1}) \odot \sigma(W_{LPCA2} \circ F_{LP} + b_{LPCA2}) \tag{6}$$

where, $F_{LPCA}$ is output feature maps of the LPCA module. $F_{LP}$ and $F_{HS}$ is the feature maps generated by HR-SinSAR and LR-PolSAR respectively. $W_{LPCA1}$ and $W_{LPCA2}$ represent the convolution kernel of the LPCA module, $b_{LPCA1}$ and $b_{LPCA2}$ represent the bias of the LPCA module.

Different from the existing channel attention mechanism, the channel attention mechanism in LPCA takes the large data dynamic range of PolSAR data and the complex data distribution into account. We remove the pooling operation that is widely used in existing channel attention to avoid numerical anomalies caused by mean pooling or maximum pooling. In the proposed channel attention module, we directly normalize the feature maps extracted by convolutional layer with the sigmoid function to obtain the channel attention weight. Subsequently, we multiply the weights with feature maps to get feature maps weighted by the channel attention mechanism. Different from the global channel attention weight of the existing channel attention mechanism, the proposed method uses local weights, which can better adapt to the characteristics of the large dynamic range of PolSAR data.

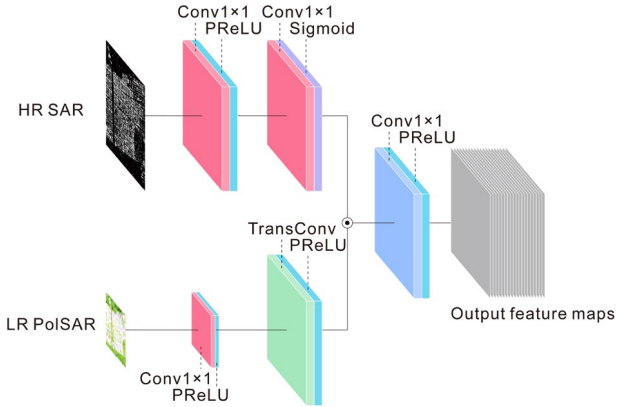 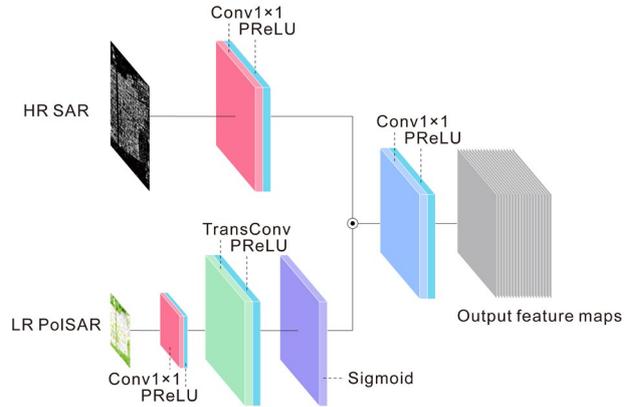

**Fig. 6.** The HSSA module.  **Fig. 7.** The LPCA module.

*2.5.3 Information Fusion Strategy of HSSA and LPCA*

After processing by the HSSA module and the LPCA module, spatially weighted feature maps and channel-weighted feature maps are obtained. To treat feature maps extracted by two modules equally, we use the cascading strategy for information fusion of the feature maps obtained by the two modules. The information fusion strategy can be written as follows:

$$F_{fusion} = Concat(F_{HSSA}, F_{LPCA}) \tag{7}$$

where, $Concat(.)$ is the cascade function, used to concatenate the feature maps generated by the HSSA module and LPCA module respectively.

### 2.6. Loss Function Based on Physical Imaging Mechanism

In terms of loss functions, we design the loss functions from two aspects: the numerical value of the PolSAR image and the physical imaging mechanism of the PolSAR image. In terms of numerical value, MSE loss is adopted to constrain numerical values of the PolSAR images. For the training set $\{C_x^i, C_y^i, I_x^i\}_{i=1}^{N}$, we use transposed convolution to get the up-sampling result $C_u^i$ of $C_y^i$, and use MSE loss for calculating loss between the residual layers $\Re^i = C_x^i - C_u^i$ and results of PSFN $f_{net}(C_y^i, I_x^i)$. Among them, $N$ represents the image pairs number. $C_x^i$ represents the HR-PolSAR image patch, and $C_y^i$ represents the LR-PolSAR image patch, $I_x^i$ represents the intensity image of HR-SinSAR image patch. The numerical loss function defined as:

$$L_{val}(\Theta) = \frac{1}{2N} \sum_{i=1}^{N} \left\| \Re^i - f_{net}(C_y^i, I_x^i) \right\|_2^2 \tag{8}$$

where $\Theta$ is the parameters of PSFN, $f_{net}(.)$ is the result of residual network.

Besides, we construct a polarimetric loss function from a physical imaging mechanism. For a given HR-PolSAR and HR-SinSAR image pair, the HR-SinSAR image can be regarded as the subset of the HR-PolSAR image in terms of polarimetric information. Therefore, for the PolSAR fusion results, the intensity image in a certain polarimetric mode of it should be consistent with the HR-SinSAR intensity image of the corresponding polarimetric mode. Therefore, considering the physical imaging relationship from the PolSAR image to the SinSAR image, the following polarimetric loss function can be used to constrain the network training process.

$$L_{phy}(\Theta) = \frac{1}{2N} \sum_{i=1}^{N} \left\| (I_x^i - \Im(C_u^i)) - \Im(f_{net}(C_y^i, I_x^i)) \right\|_2^2 \tag{9}$$

The total loss function of PSFN is expressed as:

$$L_{total} = \alpha L_{val}(\Theta) + \beta L_{phy}(\Theta) \tag{10}$$

Among them, the regularization parameters $\alpha$ and $\beta$ are adaptively determined, and they satisfy the following relationship:

$$\alpha = \frac{L_{val}(\Theta)}{L_{val}(\Theta) + L_{phy}(\Theta)}, \beta = \frac{L_{phy}(\Theta)}{L_{val}(\Theta) + L_{phy}(\Theta)} \tag{11}$$

The regularization parameters are adaptively adjusted through the numerical values of the loss functions, and higher weights are assigned to the larger loss so that the training process can converge faster.

Besides, according to the difference in polarimetric mode of input HR-SinSAR image, and under the condition of satisfying the numerical relationship of the equation (2), the polarimetric loss can be decomposed into three independent sub-loss-functions, as shown below:

$$L_{HH}(\Theta) = \frac{1}{2N} \sum_{i=1}^{N} \left\| (I_{HH}^i - \Im_{HH}(C_u^i)) - \Im_{HH}(f_{net}(C_y^i, I_{HH}^i)) \right\|_2^2 \tag{12}$$

$$L_{HV}(\Theta) = \frac{1}{2N} \sum_{i=1}^{N} \left\| \left( I_{HV}^i - \frac{\Im_{HV}(C_u^i)}{2} \right) - \frac{\Im_{HV}\left( f_{net}(C_y^i, I_{HV}^i) \right)}{2} \right\|_2^2 \tag{13}$$

$$L_{VV}(\Theta) = \frac{1}{2N} \sum_{i=1}^{N} \left\| \left( I_{VV}^i - \Im_{VV}(C_u^i) \right) - \Im_{VV}\left( f_{net}(C_y^i, I_{VV}^i) \right) \right\|_2^2 \tag{14}$$

where, $L_{HH}(.)$, $L_{HV}(.)$, and $L_{VV}(.)$ respectively represent the polarimetric loss sub-functions when using $I_{HH}$, $I_{HV}$, and $I_{VV}$ SinSAR data. $\Im_{HH}(.)$, $\Im_{HV}(.)$, $\Im_{VV}(.)$ represent the polarimetric intensity image extraction operators under polarimetric modes $HH$, $HV$, and $VV$ respectively. It should be noted that when the proposed fusion network uses a specific polarimetric mode SinSAR data to assist in the reconstruction of LR-PolSAR images, only a single sub-polarimetric loss function corresponding to the polarimetric mode is used.

## 3. EXPERIMENTAL RESULTS

### 3.1 Experimental Dataset

Datasets used in training process and test process are image pairs with four polarimetric channels PolSAR images and the auxiliary data of the SinSAR images. The data used comes from two SAR imaging systems: RADARSAT-2 and AIRSAR. Due to the limitation of the amount of available data, the data used for network training is relatively limited. Therefore, this paper only uses part of RADARSAT-2 images in San Francisco region for training, and the other is used to test. RADARSAT-2 images in Vancouver region and AIRSAR images are used in the network test process. Due to the imaging characteristics of SAR systems, the spatial resolution of PolSAR images in azimuth and range direction is usually different. In addition, the difference in resolution may confuse the reconstruction effect of the two directions in visual evaluation. Therefore, in the experiment, we processed the azimuth and range to the same resolution by performing multi-look processing on the SAR image. Besides, the radiation calibration and despeckle are performed as preprocessing by using PolSARprov5.0.

#### 3.1.1 Training Data

In terms of training data, two pairs of HR-PolSAR and HR-SinSAR image pairs are used to generate training data. The training data are all C-band data of RADARSAT-2. The train dataset details are listed in Table I. During the training process, the patch sizes of the PolSAR image and the SinSAR intensity image are $40 \times 40 \times 9$ and $40 \times 40 \times 1$, respectively. Besides, we used the PolSARpro V5.0 software provided by ESA to generate LR-PolSAR data with the patch size of $20 \times 20 \times 9$. In total, 71040 patches were randomly generated and used for network training.

**TABLE I** Train dataset information

| Sensor | Band | Region | Looks | Pixel spacing | Size | mode |
|---|---|---|---|---|---|---|
| RADARSAT-2 | C | San Francisco | 1 | 8 m | 6800×2400×9 | HR full-polarimetric |
| | C | San Francisco | 1 | 8 m | 6800×2400×1 | HR single-polarization |
| RADARSAT-2 | C | San Francisco | 1 | 8 m | 5200×2400×9 | HR full-polarimetric |
| | C | San Francisco | 1 | 8 m | 5200×2400×1 | HR single-polarization |

*3.1.2 Test Data*

In terms of test data, three pairs of LR-PolSAR and HR-SinSAR image pairs are used to test the network. The test data includes two scene image pairs of RADARSAT-2, and one scene image pairs of AIRSAR. The test dataset details are listed in Table II.

**TABLE II** Test datasets information

| Sensor | Band | Region | Looks | Nominal resolution | Size | mode |
|---|---|---|---|---|---|---|
| RADARSAT-2 | C | San Francisco | 1 | 16 m | 1200×1200×9 | LR fully polarimetric |
| | C | San Francisco | 1 | 8 m | 2400×2400×1 | HR single-polarization |
| RADARSAT-2 | C | Vancouver | 1 | 16 m | 3200×1600×9 | LR fully polarimetric |
| | C | Vancouver | 1 | 8 m | 6400×3200×1 | HR single-polarization |
| AIRSAR | L | San Francisco | 4 | 24m | 450×512×9 | LR fully polarimetric |
| | L | San Francisco | 4 | 12m | 900×1024×1 | HR single-polarization |

In the auxiliary data polarimetric mode selection, we compared the intensity images of RADARSAT-2 images in HH, VV, and HV polarimetric modes. As shown in Fig. 8, compared to the co-polarized channel, the intensity image of HV is only stronger in urban built-up areas, with lower values and weaker features in most areas. This phenomenon may make it difficult for us to extract features when extracting information from HV channel intensity images. Also, we noticed that the urban built-up area in the HH channel intensity image has a strong contrast with other areas. As auxiliary information, it may lead to excessive enhancement of urban built-up areas and weakening of other areas. Therefore, in the experiment of this paper, we use the intensity image of the co-polarized channel VV with a more balanced intensity as input to ensure that the spatial information of the SinSAR image can be better utilized.

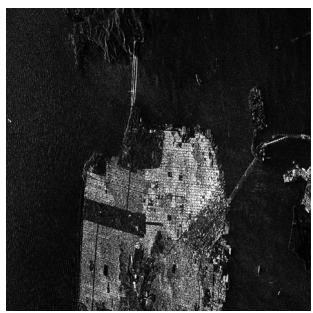 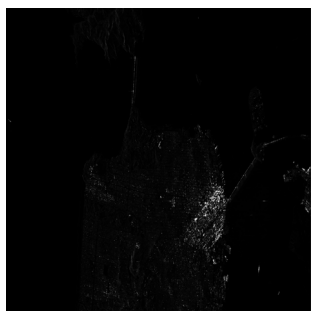 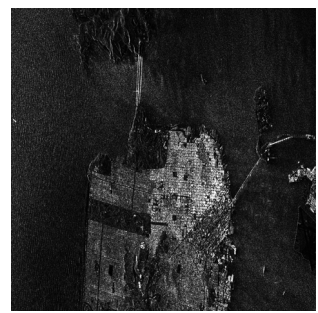

HH polarimetric mode　　　　　　　　HV polarimetric mode　　　　　　　　VV polarimetric mode

**Fig. 8.** Intensity images in different polarimetric modes.

*3.2 Parameter Settings*

The proposed network includes four modules, LPSR, HSFE, HSSA, and LPCA. In Table III, we list the detailed configuration information of each module of the network. We use Kaiming normal[40] to initialize the network model parameters. The Adam[41] is used to update weights. There are 100 epochs for network training. The initial learning rate is 0.0001, and it drops by half every 50 epochs. We use the PyTorch framework to train the fusion network in the Win10 environment, and an NVIDIA Quadro P4000 GPU.

**TABLE III** The Configuration of PSFN.

| Module | Layer | Configuration |
|---|---|---|
| LPSR | Complex structure block | See the Complex structure block configuration below for details |
|  | Transposed convolutional | TransConv (64×4×4, stride=2, pad=1) + PReLU |
|  | ResBlock | See the ResBlock configuration below for details |
| HSFE | Convolutional | Conv1 (64×3×3, stride=1, pad=1) + PReLU: |
|  | ResBlock | See the ResBlock configuration below for details |
|  | Spatial attention block | See the Spatial attention block configuration below for details |
|  | Convolutional | Conv2 (64×3×3, stride=1, pad=1) + PReLU: |
| HSSA | Convolutional | Conv1 (64×3×3, stride=1, pad=1) + PReLU |
|  | Transposed convolutional | TransConv (64×4×4, stride=2, pad=1) + PReLU |
|  | Convolutional | Conv2 (64×3×3, stride=1, pad=1) + PReLU + Sigmoid |
|  | Convolutional | Conv3 (64×3×3, stride=1, pad=1) + PReLU: |
| LPCA | Convolutional | Conv1 (64×3×3, stride=1, pad=1) + PReLU |
|  | Transposed convolutional | TransConv (64×4×4, stride=2, pad=1) + PReLU + Sigmoid |
|  | Convolutional | Conv2 (64×3×3, stride=1, pad=1) + PReLU: |
|  | Convolutional | Conv3 (64×3×3, stride=1, pad=1) + PReLU: |
| ResBlock | Convolutional | Conv1 (128×3×3, stride=1, pad=1) + PReLU |
|  | Convolutional | Conv2 (64×3×3, stride=1, pad=1) |
| Spatial attention block | Convolutional | Conv1 (64×3×3, stride=1, pad=1) + PReLU |
|  | Convolutional | Conv2 (1×3×3, stride=1, pad=1) + Sigmoid |
|  | Convolutional | Conv3 (64×3×3, stride=1, pad=1) + PReLU |

*3.3 Comparison methods and Quantitative Evaluation Measures*

We conducted quantitative evaluation experiments on various datasets to confirm the effectiveness of PSFN. Four PolSAR image reconstruction algorithms were selected for comparison, including Bicubic interpolation[42], SRPSC[14], MSSR[23], PSSR[24]. Both MSSR and PSSR are retrained using the data set used in this study.

In the experimental analysis part, to intuitively reflect the physical mechanism corresponding to the research target in the PolSAR image rather than simply the degree of numerical retention, we carried out a data conversion on the reconstruction results of the PolSAR image, and transformed the $C_3$ matrix into a Pauli coherency matrix for subsequent quantitative evaluation experiments through the coherency matrix and covariance conversion equation[43]–[45], as shown in Equation (15) and Equation (16). Subsequently, we perform Pauli decomposition of the coherency matrix to obtain three polarimetric components, as shown in the Equation (17). Then calculate the quantitative index of the polarimetric component.

$$T_3 = U_{3(L \to P)} C_3 U_{3(L \to P)}^{-1} \tag{15}$$

$$U_{3(L \to P)} = \frac{1}{\sqrt{2}} \begin{bmatrix} 1 & 0 & 1 \\ 1 & 0 & -1 \\ 0 & \sqrt{2} & 0 \end{bmatrix} \tag{16}$$

$$P_1 = \frac{S_{HH} + S_{VV}}{\sqrt{2}}, P_2 = \frac{S_{HH} - S_{VV}}{\sqrt{2}}, P_3 = \frac{2S_{HV}}{\sqrt{2}} \tag{17}$$

where, $P_1, P_2, P_3$ represents the odd scattering mechanism, the double scattering mechanism, and the volume scattering mechanism respectively.

In the experimental analysis, PSNR and MAE are employed to quantitatively assess the fusion result. Among them, the higher the PSNR value, the higher the image quality. The smaller the MAE, the more similar the predicted results are to the ground truth. We calculated PSNR and MAE in polarimetric components of Pauli-decomposition to evaluate spatial information reconstruction ability and polarimetric information retention ability of the algorithm.

### 3.4 Simulated Experiments

We analyze the experimental results from the following three aspects to verify the effectiveness of PSFN in reconstructing the LR-PolSAR image resolution.

#### 3.4.1 Simulated experiments of RADARSAT-2 in San Francisco area

This set of experiments is used to verify the reconstruction effect of the fusion network under the same satellite and similar imaging environments, such as roughly the same imaging area, and similar imaging angles. This set of experiments is mainly aimed at the application scenario of using the same sensor to monitor the same area for a long time series.

**TABLE IV** Quantitative Evaluation Results of RADARSAT-2 (San Francisco).

| Method | Bicubic | SRPSC | MSSR | PSSR | PSFN |
|---|---|---|---|---|---|
| PSNR ($|P_1|^2$) | 45.63 | 46.04 | 46.08 | <u>47.25</u> | **52.04** |

| | | | | | |
|---|---|---|---|---|---|
| PSNR ($|P_2|^2$) | 43.16 | 42.84 | 43.19 | <u>43.42</u> | **48.82** |
| PSNR ($|P_3|^2$) | 51.39 | 50.89 | 52.23 | <u>52.53</u> | **53.26** |
| PSNR (mean) | 46.73 | 46.59 | 47.17 | <u>47.73</u> | **51.37** |
| MAE ($|P_1|^2$) | 0.28 | <u>0.24</u> | 0.36 | 0.25 | **0.08** |
| MAE ($|P_2|^2$) | 0.20 | <u>0.19</u> | 0.25 | 0.21 | **0.10** |
| MAE ($|P_3|^2$) | 0.03 | <u>0.03</u> | 0.04 | 0.04 | **0.03** |
| MAE (mean) | 0.17 | <u>0.15</u> | 0.22 | 0.17 | **0.07** |

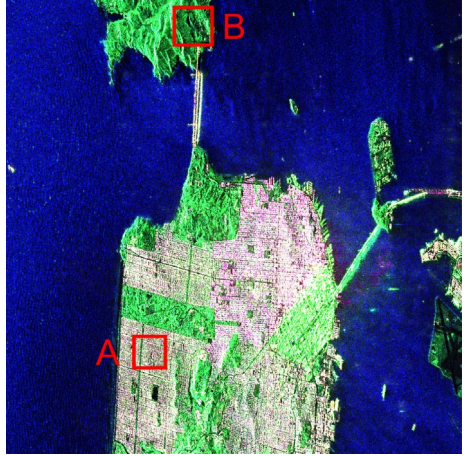

**Fig. 9.** The PSFN fusion result for the RADARSAT-2 imagery (San Francisco).

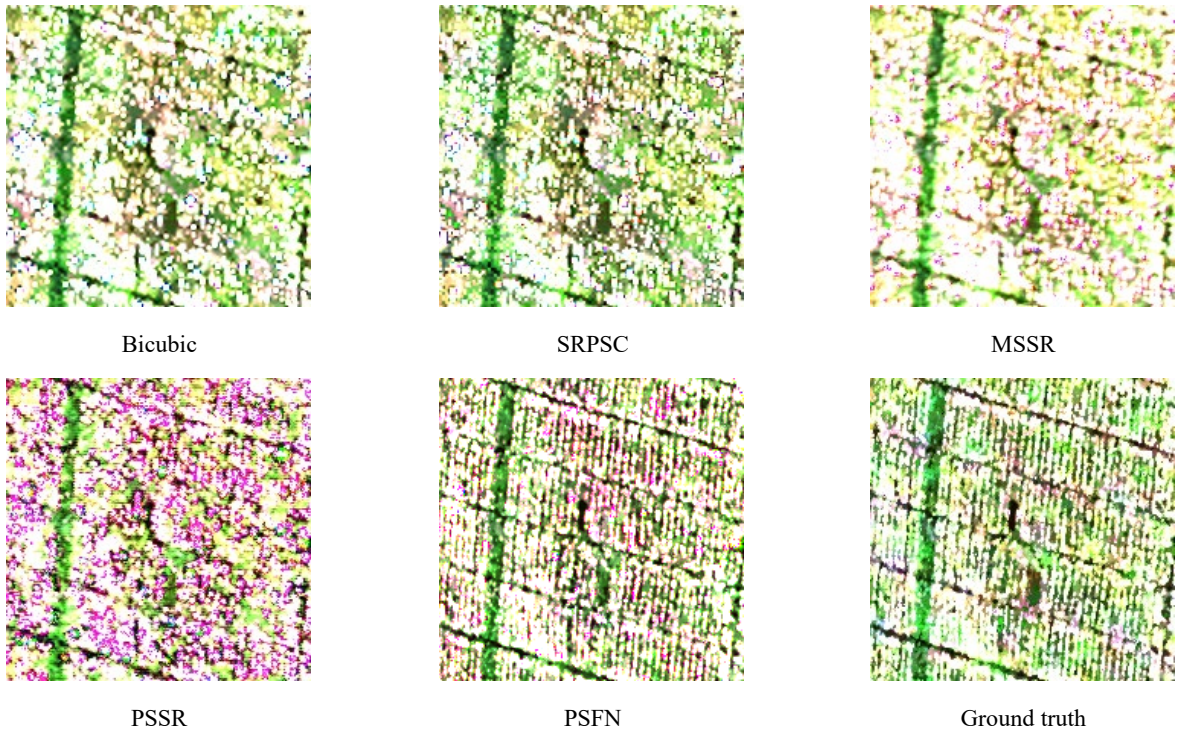

**Fig. 10.** Comparison results in the urban area.

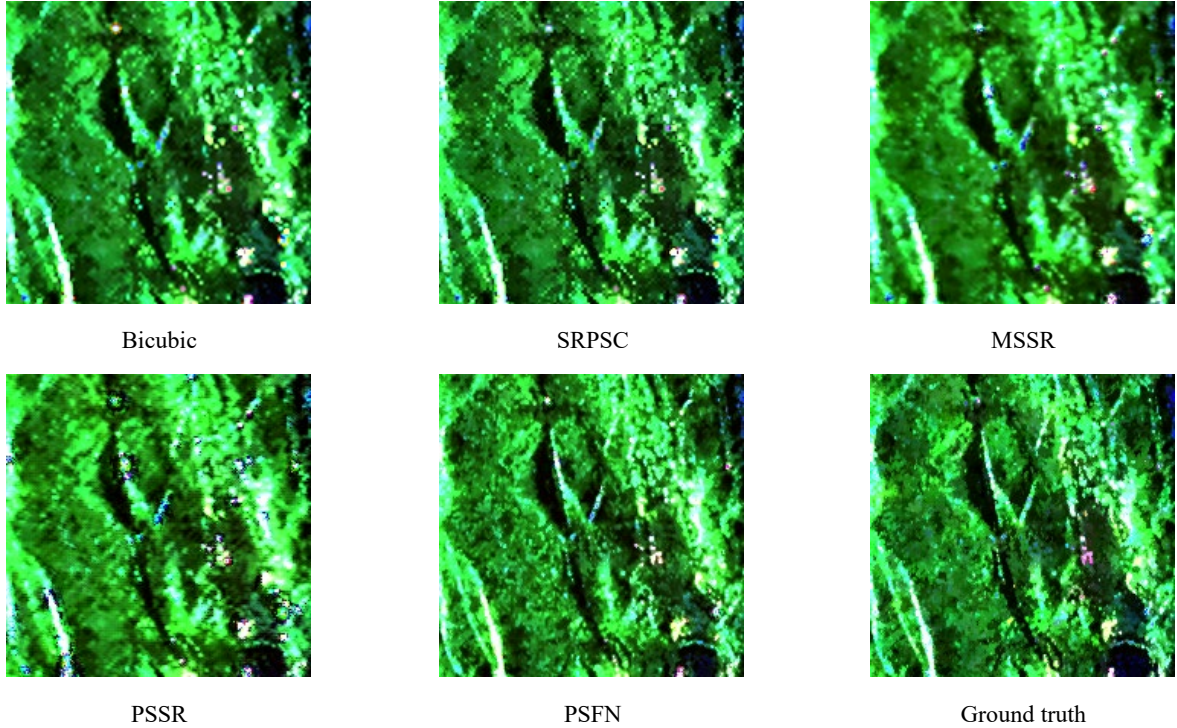

| Bicubic | SRPSC | MSSR |
| PSSR | PSFN | Ground truth |

**Fig. 11.** Comparison results in the vegetation area.

This set of experiments included a scene RADARSAT-2 image in San Francisco region which is displayed in Fig. 9. We selected two typical targets for visual evaluation, including urban areas and vegetation areas, as shown in Fig. 10 and Fig. 11. Compared with the deep-learning-based method, the results of bicubic interpolation are too smooth, while the results of the SRPSC method have an obvious grid effect. The results of the MSSR method have color distortion and are relatively smooth. The result of PSSR has some artifacts in some areas. The proposed fusion network has a better ability to maintain texture information, and the color of the Pauli-composite image is similar to ground truth. In the urban area, the proposed method keeps the shape of the buildings better. In the vegetation area, the result of PSFN is sharper at the edge and the ridgeline is also clearer. In Table IV, the quantitative evaluation results illustrate that the proposed fusion network has a significant improvement in PSNR and MAE. It indicates that the results of the fusion network more resemble ground truth in texture information and have a lower absolute error than the comparison algorithm.

*3.4.2 Simulated experiments of RADARSAT-2 in Vancouver area*

This set of experiments is mainly used to verify the effectiveness of the fusion network under an inconsistent imaging environment, which is aimed at the main application scenario of this paper. That is, the fusion of PolSAR images acquired by a certain sensor under different imaging environments, such as the inconsistent imaging areas and inconsistent orbits.

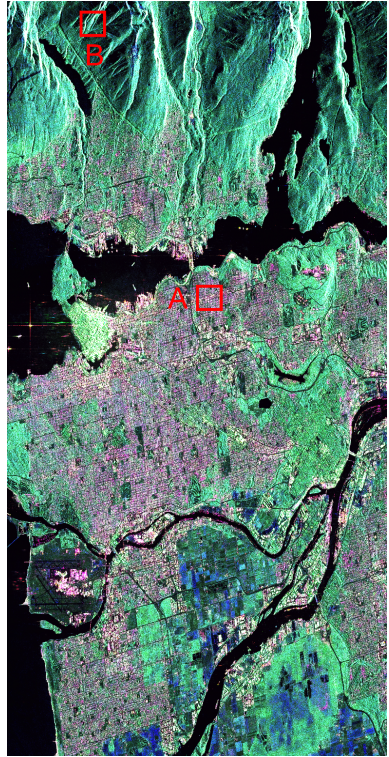

**Fig. 12.** The PSFN fusion result for the RADARSAT-2 imagery (Vancouver).

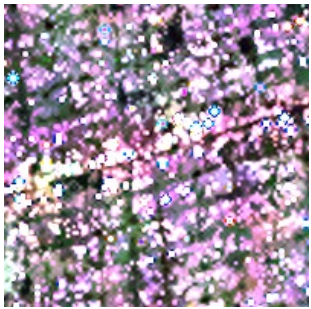

Bicubic

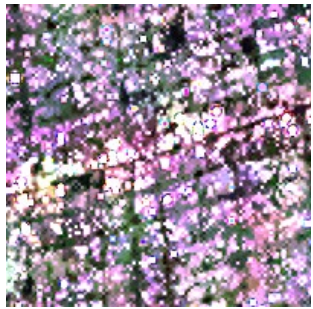

SRPSC

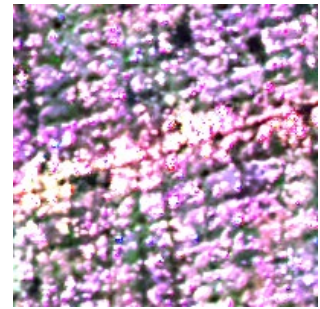

MSSR

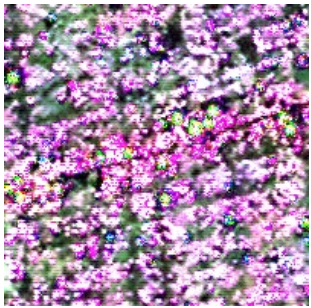

PSSR

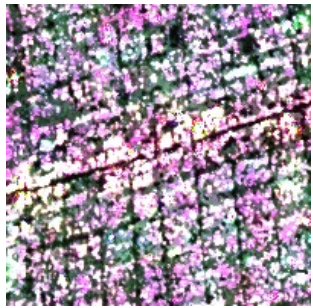

PSFN

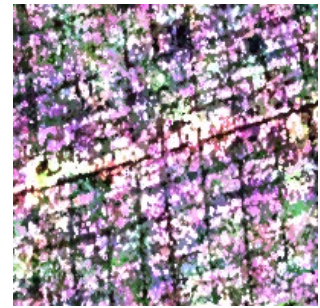

Ground truth

**Fig. 13.** Comparison results in the urban area.

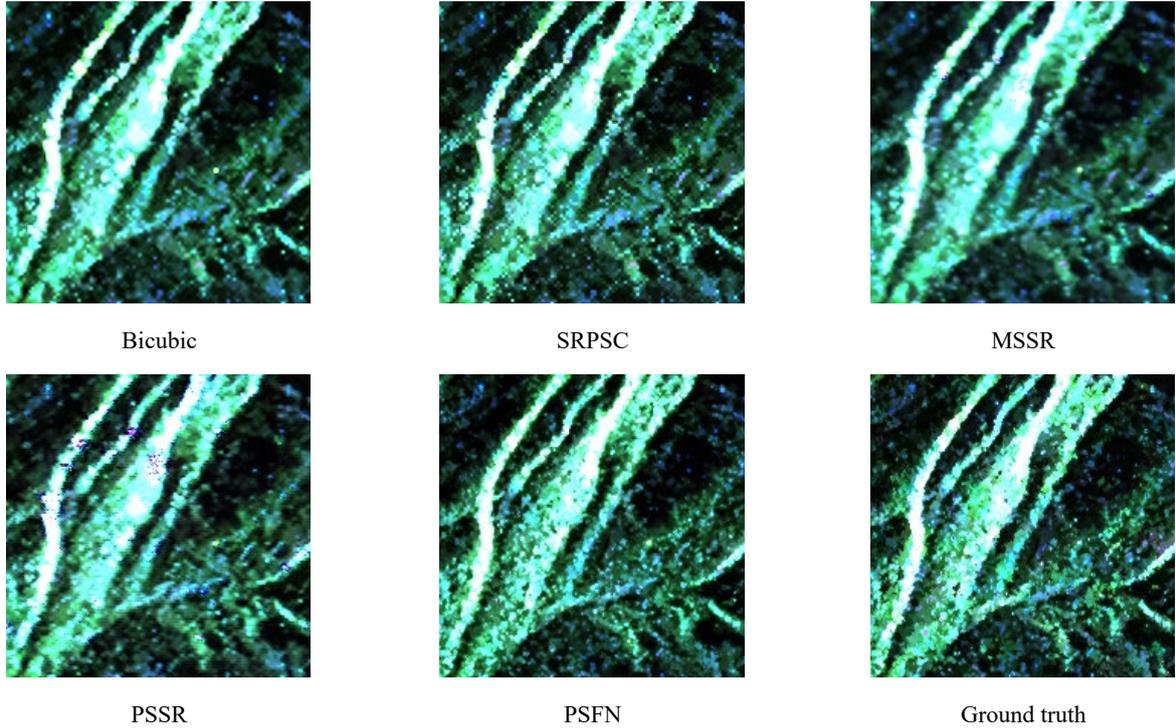

| Bicubic | SRPSC | MSSR |
| PSSR | PSFN | Ground truth |

**Fig. 14.** Comparison results in the vegetation area.

This set of experiments included a scene RADARSAT-2 data in the Vancouver area which is displayed in Fig. 12. In this set of experiments, as shown in Fig. 13 and Fig. 14, urban densely built-up areas and vegetation areas were selected for visual evaluation. Compared with the deep-learning-based method, the interpolation result is too smooth, and the result of the SRPSC method is mosaic-like due to the grid effect. The target boundary of the reconstruction result of the MSSR method is relatively smooth, and the PSSR method introduces obvious artificial artifacts while reconstructing the details. The proposed method can effectively reconstruct texture details, and the boundaries between targets are clear and sharp. In the urban densely built-up area, the proposed method can effectively reconstruct the vertical road, but the comparison algorithm fails to reconstruct the vertical road. Compared with other algorithms, the proposed method has richer texture details and sharper ridgelines in vegetation areas. In Table V, the proposed fusion network has greatly improved in PSNR and MAE over comparison algorithm.

**TABLE V** Quantitative Evaluation Results of RADARSAT-2 (Vancouver).

| Method | Bicubic | SRPSC | MSSR | PSSR | PSFN |
| --- | --- | --- | --- | --- | --- |
| PSNR ($|P_1|^2$) | 46.32 | 45.84 | 47.32 | <u>47.33</u> | **53.08** |
| PSNR ($|P_2|^2$) | 43.49 | 43.04 | 44.40 | <u>44.55</u> | **51.08** |
| PSNR ($|P_3|^2$) | 53.63 | 53.31 | 55.03 | <u>54.98</u> | **56.49** |
| PSNR (mean) | 47.81 | 47.40 | 48.92 | <u>48.95</u> | **53.55** |
| MAE ($|P_1|^2$) | 0.17 | 0.16 | 0.16 | <u>0.15</u> | **0.07** |

| | | | | | |
|---|---|---|---|---|---|
| MAE ($|P_2|^2$) | 0.15 | 0.14 | 0.16 | <u>0.13</u> | **0.07** |
| MAE ($|P_3|^2$) | 0.04 | 0.04 | 0.04 | <u>0.04</u> | **0.03** |
| MAE (mean) | 0.12 | 0.11 | 0.12 | <u>0.11</u> | **0.06** |

*3.4.3 Simulated experiments of AIRSAR in San Francisco area*

This set of experiments is mainly used to verify the fusion network generalization ability and reconstruction ability under different sensors and inconsistent imaging environments. It included a scene AIRSAR data, as displayed in Fig. 15. We selected urban areas, vegetation and sea areas for visual evaluation.

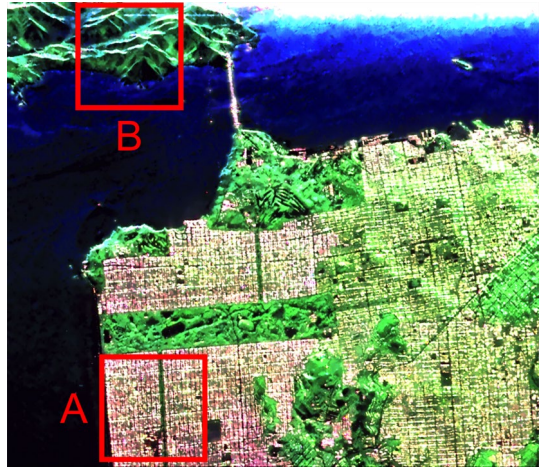

**Fig. 15.** The PSFN fusion result for the AIRSAR imagery (San Francisco).

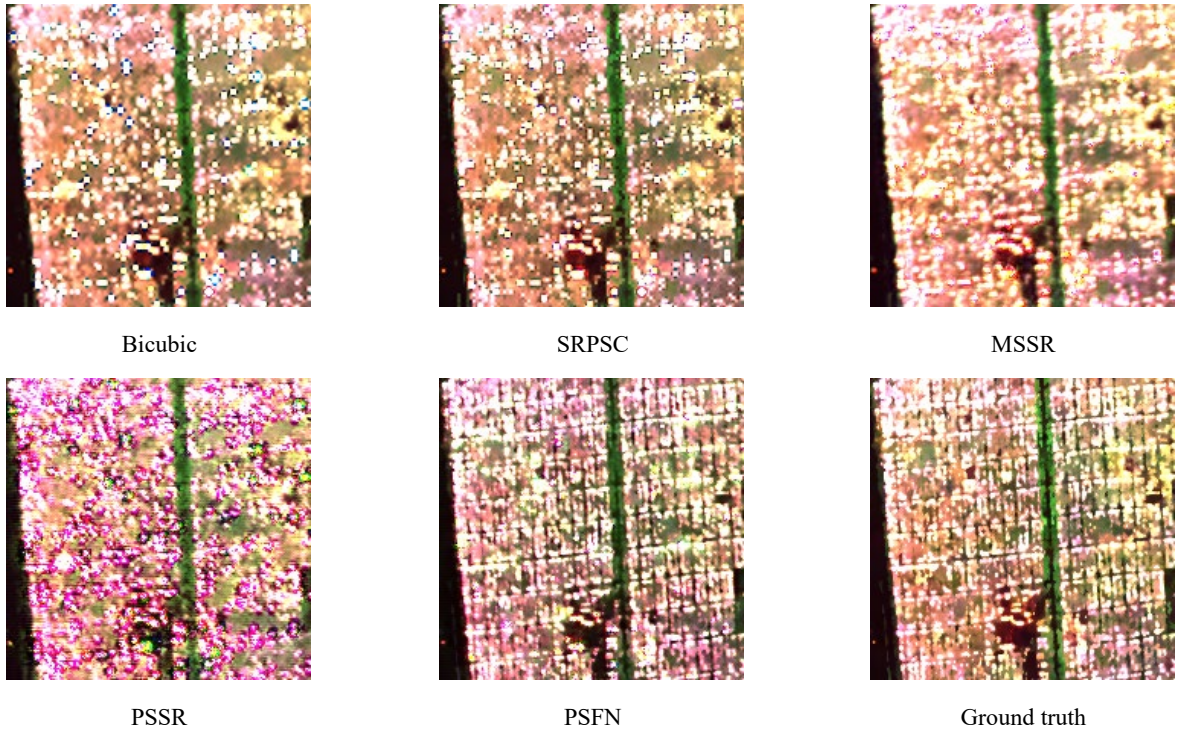

**Fig. 16.** Comparison results in the urban area.

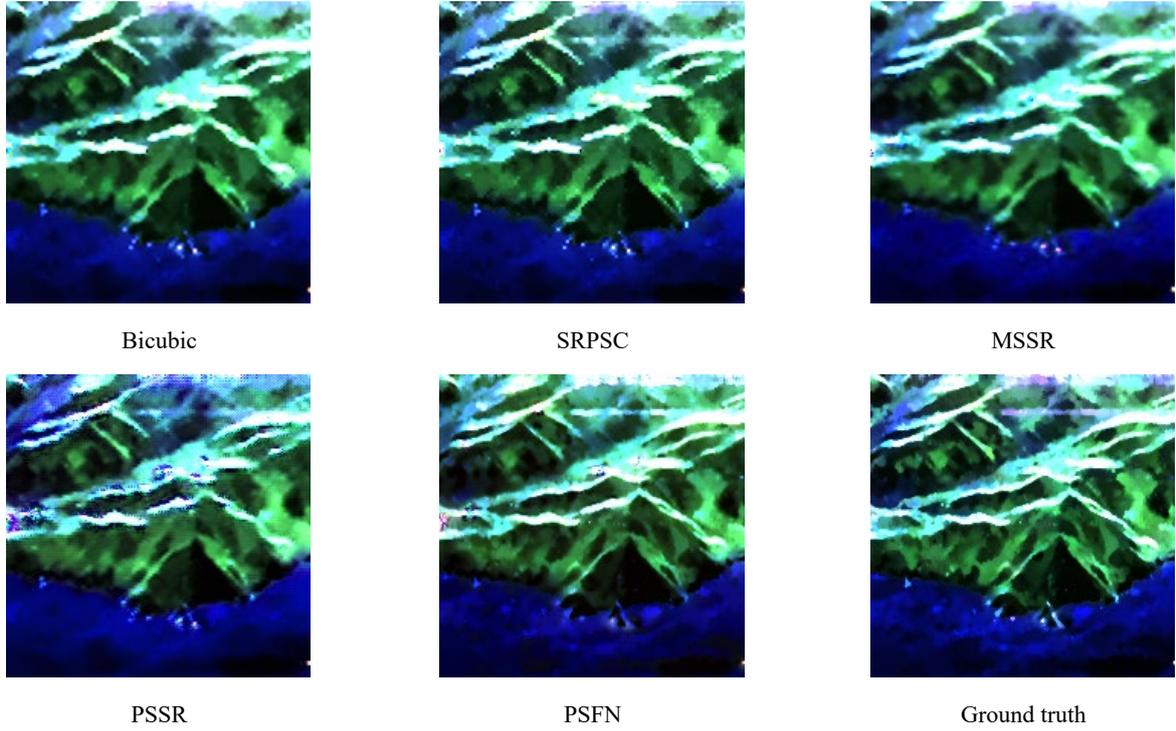

|  | Bicubic | SRPSC | MSSR | PSSR | PSFN | Ground truth |

Fig. 17. Comparison results in the vegetation and sea area.

As shown in Fig. 16 and Fig. 17, the color of the reconstruction result of the traditional method is maintained well, but the spatial detail reconstruction is insufficient, which has an obvious grid effect. Compared with the traditional method, the MSSR method eliminates the grid effect, but the spatial details are too smooth. The PSSR method has obvious artifacts. The proposed fusion network has a significant improvement in spatial information, and texture information of the result is consistent with ground truth. However, we noticed that the proposed method has color distortion in parts of urban areas, which may be caused by the large difference in the numerical dynamic range between different sensors and the lack of AIRSAR data during the training process. The quantitative evaluation is summarized in Table VI. the result of the PSFN has significantly improved in PSNR and MAE over comparison algorithm, which indicates the better generalization ability and stronger robustness of the proposed fusion network.

**TABLE VI** Quantitative Evaluation Results of AIRSAR (San Francisco).

| Method | Bicubic | SRPSC | MSSR | PSSR | PSFN |
|---|---|---|---|---|---|
| PSNR ($|P_1|^2$) | 50.29 | 51.44 | 47.72 | <u>53.93</u> | **62.15** |
| PSNR ($|P_2|^2$) | 44.94 | 45.62 | 46.08 | <u>46.49</u> | **58.33** |
| PSNR ($|P_3|^2$) | 51.09 | 51.58 | 51.49 | <u>52.00</u> | **54.35** |
| PSNR (mean) | 48.78 | 49.55 | 48.43 | <u>50.81</u> | **58.28** |
| MAE ($|P_1|^2$) | 0.16 | <u>0.09</u> | 0.36 | 0.15 | **0.04** |
| MAE ($|P_2|^2$) | 0.32 | <u>0.23</u> | 0.38 | 0.25 | **0.04** |
| MAE ($|P_3|^2$) | 0.11 | <u>0.08</u> | 0.18 | 0.12 | **0.06** |

|  | MAE (mean) | 0.20 | <u>0.13</u> | 0.31 | 0.18 | **0.04** |

### 3.5 Verification of Real Experiments

To confirm the effect of PSFN in practical application, the fine mode SAR image and standard mode SAR image of RADARSAT-2 are adopted in real experiments. The details of the two types of SAR images are listed in Table VII.

**TABLE VII** Data information of real experiments.

| Sensor | Band | Region | Looks | Nominal resolution | mode |
|---|---|---|---|---|---|
| RADARSAT-2 | C | Quebec | 1 | 8m | Fine Quad-Pol |
| RADARSAT-2 | C | Quebec | 1 | 25m | Standard Quad-Pol |

We performed radiometric calibration and despeckle processing on the PolSAR images, and performed the necessary registration processing on the two PolSAR images. Besides, since the azimuth resolution of the images obtained in the fine mode and the standard mode is the same, to more intuitively demonstrate the effect of the proposed method, we have performed multi-look processing on the azimuth resolution of the standard mode data, making $r_{azimuth} = r_{range}$.

In real experiments, we fused the LR-PolSAR image in the standard mode with the HR-SinSAR in the fine mode, and calculated the quantitative index of the fusion result with the HR-PolSAR image in the fine mode. The PolSAR image obtained in the fine mode, the PolSAR image in the standard mode, and the fusion results are listed in Fig. 18. Two typical land covers are selected for visual evaluation, including densely built-up areas and vegetation areas.

**TABLE VIII** Quantitative Evaluation Results of RADARSAT-2 (Quebec).

| Method | Bicubic | SRPSC | MSSR | PSSR | PSFN |
|---|---|---|---|---|---|
| PSNR ($|P_1|^2$) | 50.77 | 50.74 | 50.90 | <u>51.72</u> | **56.29** |
| PSNR ($|P_2|^2$) | 49.95 | 50.21 | 50.27 | <u>50.76</u> | **58.39** |
| PSNR ($|P_3|^2$) | 50.35 | 50.60 | <u>54.28</u> | 54.17 | **56.59** |
| PSNR (mean) | 50.36 | 50.52 | 51.82 | <u>52.22</u> | **57.09** |
| MAE ($|P_1|^2$) | 0.10 | <u>0.08</u> | 0.12 | 0.10 | **0.03** |
| MAE ($|P_2|^2$) | 0.10 | <u>0.09</u> | 0.14 | 0.11 | **0.02** |
| MAE ($|P_3|^2$) | 0.15 | 0.13 | 0.09 | <u>0.08</u> | **0.05** |
| MAE (mean) | 0.12 | 0.10 | 0.12 | <u>0.10</u> | **0.04** |

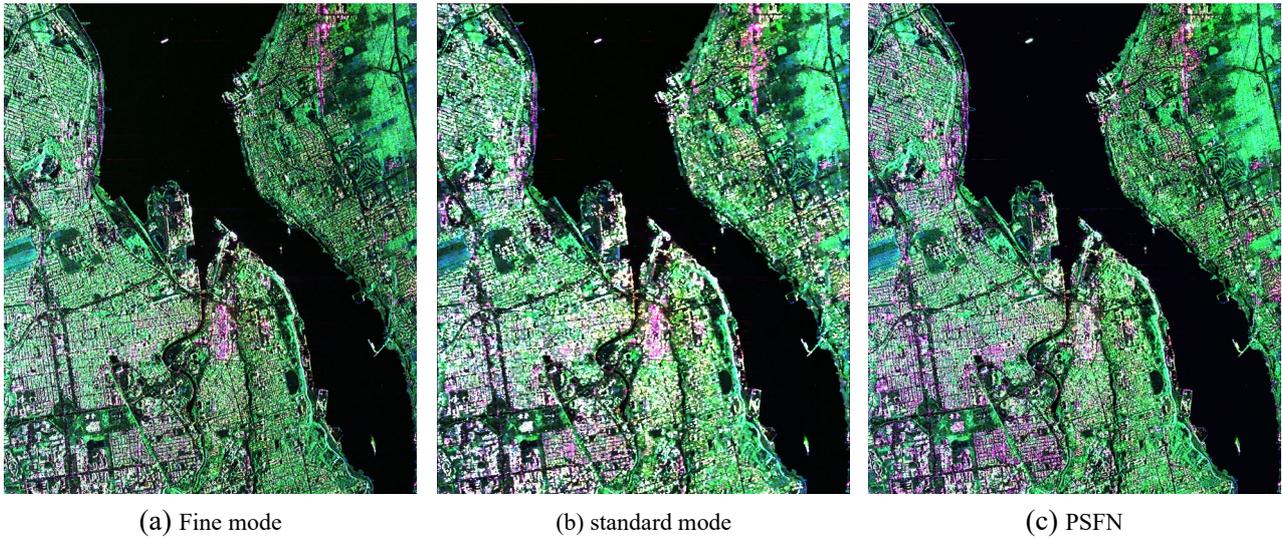

(a) Fine mode          (b) standard mode          (c) PSFN

**Fig. 18.** PolSAR image in Quebec area.

In Fig. 19, in densely built-up areas, the existing methods have poor detail information reconstruction, while the proposed method can better integrate the spatial detail information of HR-SinSAR images. In terms of color fidelity, the proposed method is more inclined to maintain the color of the data acquired in the standard mode. The reason is that in the network structure design, more consideration is given to maintaining the polarimetric information of the LR-PolSAR images rather than the polarization information of the single-polarization images. As shown in Fig. 20, in the vegetation area, the existing methods are relatively vague, and the detailed information reconstruction is insufficient. The texture details of PSFN are more consistent with the PolSAR image in fine mode, and the color is more consistent with the PolSAR image in the standard mode. In terms of quantitative assessment, the proposed method improves average PSNR by more than 4.8db, and the average MAE decreases by more than 0.06 over existing methods. Visual evaluation and quantitative results illustrate that PSFN can effectively improve spatial resolution of LR-PolSAR images in practical applications.

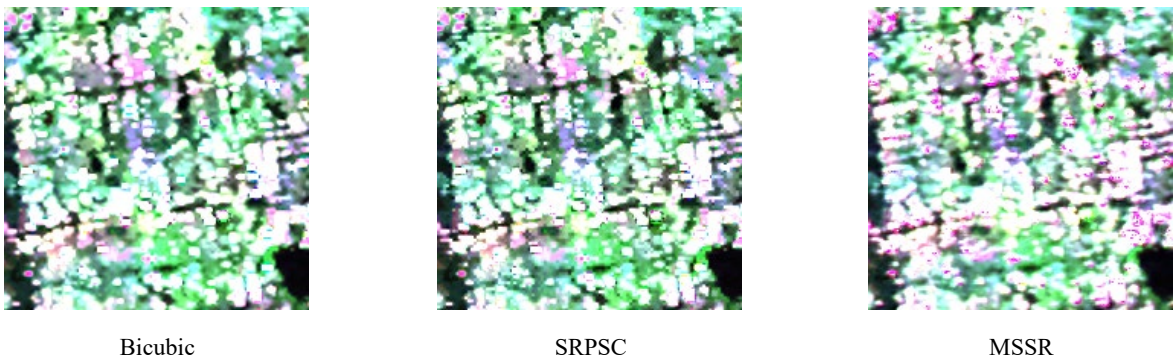

Bicubic          SRPSC          MSSR

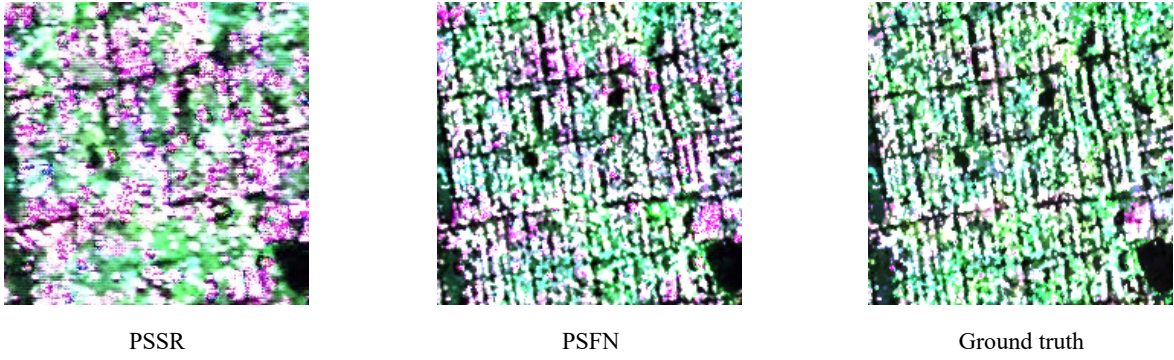

| PSSR | PSFN | Ground truth |

**Fig. 19.** Comparison results in the densely built-up areas.

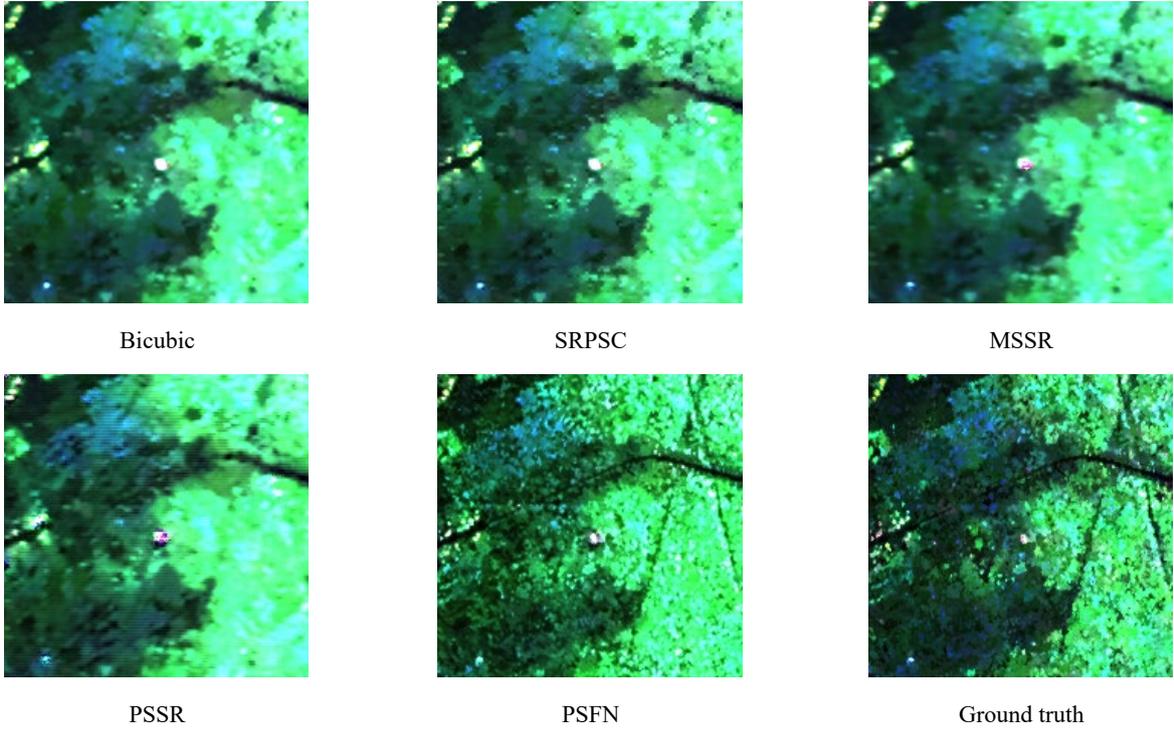

| Bicubic | SRPSC | MSSR |

| PSSR | PSFN | Ground truth |

**Fig. 20.** Comparison results in the vegetation areas.

## 4. POLARIMETRIC ANALYSIS

### 4.1. Analysis of Trace Moment ENL

In order to check if the proposed method introduces artifacts in the fusion process, the equivalent number of looks (ENL) of PolSAR image is calculated by using the covariance or coherency matrix. The ENL is expressed as follows:

$$ENL = \frac{Tr(\langle T_3 \rangle)^2}{\langle Tr(T_3 T_3) \rangle - Tr(\langle T_3 \rangle \langle T_3 \rangle)} \tag{18}$$

where $Tr(.)$ is the trace operator. $T_3$ is the PolSAR coherency matrices, $\langle . \rangle$ is neighborhood averaging, in practical applications, it is generally recommended to use the 7×7 neighborhood to calculate ENL[46], [47].

The ENL is commonly adopted to check the speckle level of SAR image. A high ENL value indicates a better suppression of speckles, meaning that the homogeneous area is smoother. The PolSAR fusion result usually cannot reconstruct the detailed information that is completely consistent with the real HR-PolSAR image, and it is often smoother than the real HR-PolSAR image. Therefore, compared to ground truth, the ENL value of fusion results are often higher. If ENL of a certain area in the fusion result is lower than that of ground truth, it indicates that the area has artificial artifacts.

We performed ENL calculations on the fusion results of RADARSAT-2 in San Francisco region and mapped results. As shown in Fig. 21, the ENL map value of PSFN is higher, indicating that the result of PSFN is smoother than ground truth, reflecting that no obvious artifacts are introduced in the reconstruction result of PSFN as a whole.

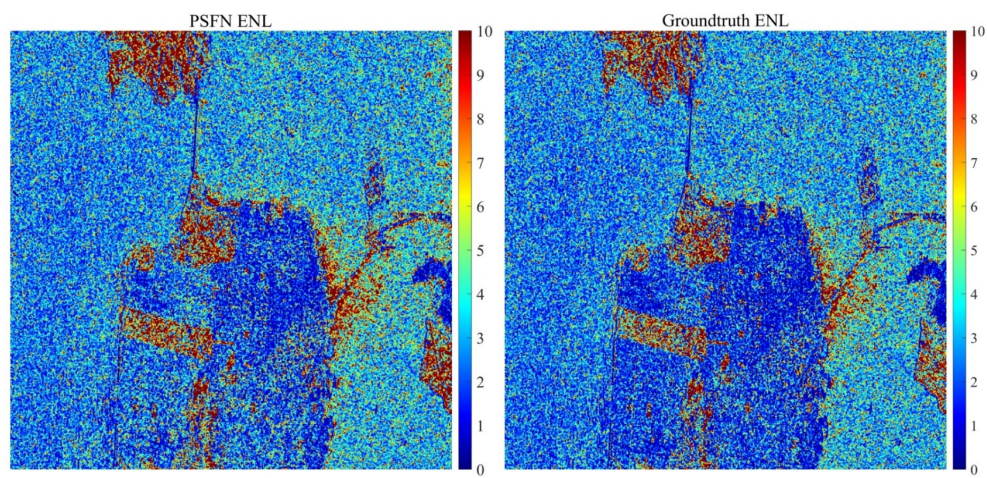

**Fig. 21.** ENL map results.

*4.2. Analysis of Polarimetric Decomposition*

To confirm the polarimetric information retention ability, we performed polarimetric target decomposition on the reconstruction results to analyze the similarities and differences between the polarimetric decomposition results of the reconstructed image and the polarimetric decomposition results of ground truth. Polarimetric decomposition is a theory that uses PolSAR polarimetric information to reveal the physical mechanism of scatters. By comparing the scattering mechanism of fusion result with that of ground truth under different scatters, it can reflect the degree of preservation of polarimetric information by the proposed method. In this paper, the Yamaguchi decomposition (Y4R)[48] is used for polarimetric decomposition of fusion results, decomposed into single scattering, double scattering, volume scattering, helix scattering.

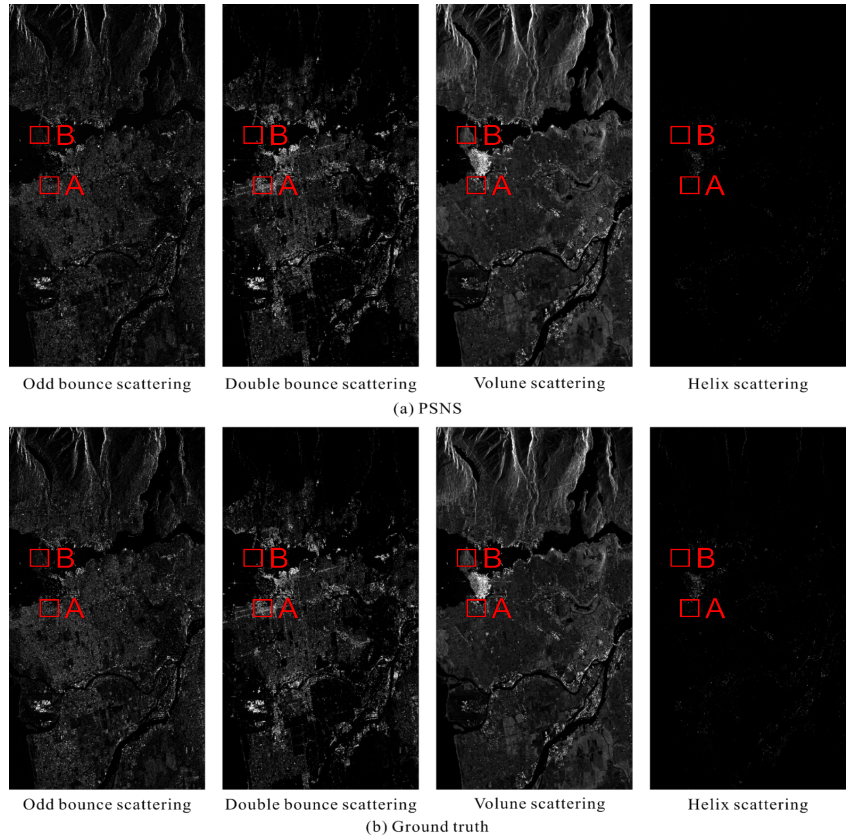

**Fig. 22.** Polarimetric decomposition results

In RADARSAT-2 PolSAR image in the Vancouver area, we selected two typical types of land cover for evaluation, including densely built-up urban areas and vegetation areas. In the urban densely built-up area, the wall of the building is perpendicular to the ground to form a dihedral angle. Therefore, double scattering is the dominant component of the scattering mechanism in this area. The scattering mechanism of densely built-up areas in urban is relatively complicated. As shown in Fig. 23, the echo signal has multiple scattering components in densely built-up areas, and single scattering and volume scattering also have a certain intensity in this area. Besides, compared to other areas, the intensity of helix scattering in this area is higher. In the vegetation area, under the action of the vegetation canopy, the scattering mechanism is mainly volume scattering. The intensity of single scattering is low, and there is almost no double scattering and helix scattering. As display in Fig. 24, polarimetric decomposition results of the fusion results in the vegetation area are consistent with that of corresponding ground truth. Polarimetric decomposition results of the fusion network in the two typical surface coverage conform to the corresponding scattering mechanism, and at the same time, its intensity is similar to the intensity of ground truth's, illustrating that the proposed fusion network can better maintain the corresponding surface coverage types in the PolSAR image polarimetric information.

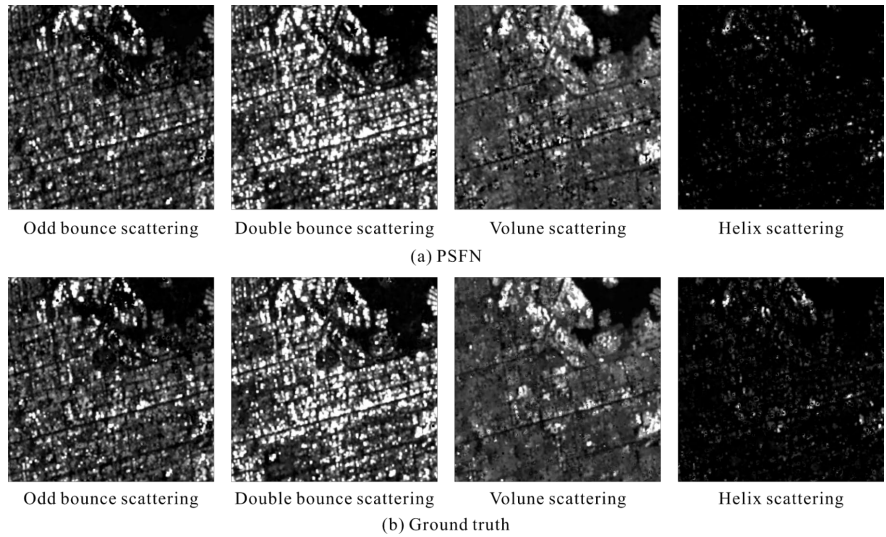

**Fig. 23.** Polarimetric decomposition results in the urban area.

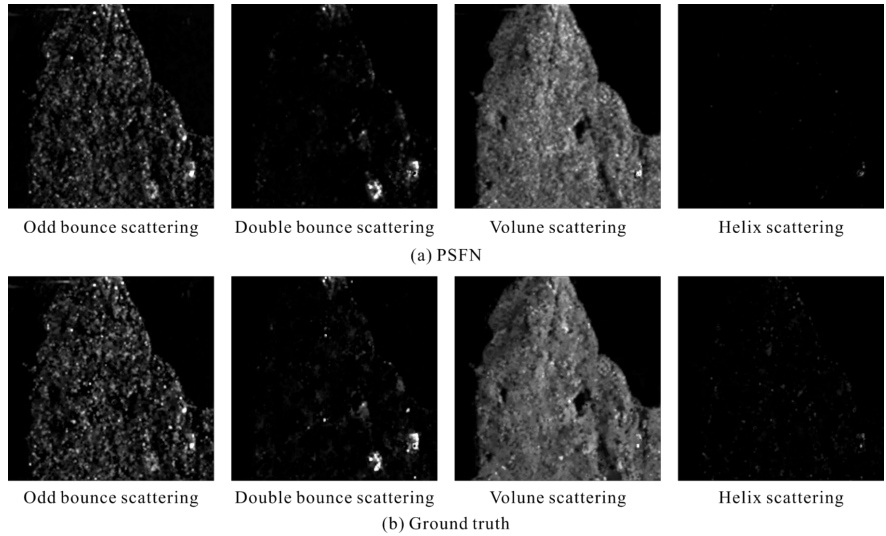

**Fig. 24.** Polarimetric decomposition results in the vegetation area.

*4.3. Analysis of Polarimetric Signature*

To analyze the extent to which the reconstruction results maintain the polarimetric characteristics, we selected two ground targets with different scattering characteristics from the reconstruction results of the RADARSAT-2 in the Quebec area and generated the polarimetric signature, including vegetation areas and urban built-up areas. Polarimetric signature is a curved surface describing the scattering characteristics of ground scatterers in any polarimetric state, which is composed of a co-polarimetric signature and a cross-polarimetric signature. To a certain extent, it reflects the change in the echo power of a target under a specific polarimetric combination.

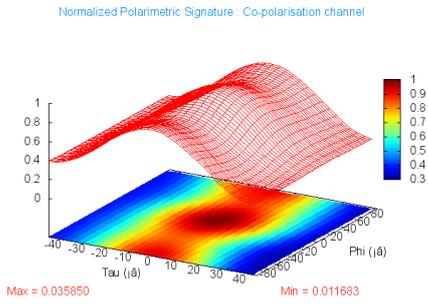

**(a)** Co-polarimetric signature of PSFN.

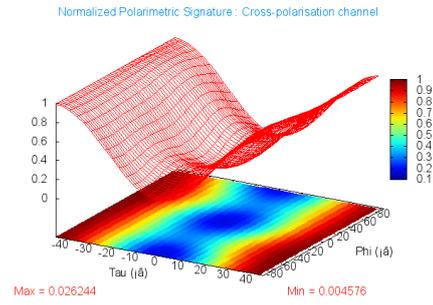

**(b)** Cross-polarimetric signature of PSFN.

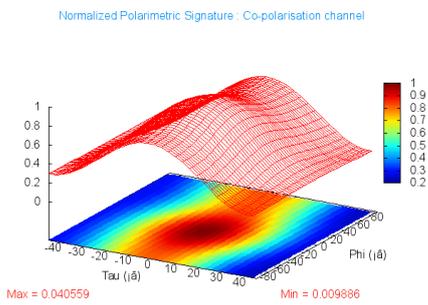

**(c)** Co-polarimetric signature of fine mode.

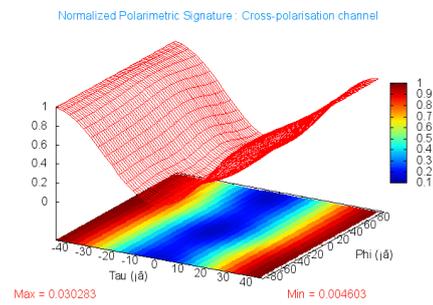

**(d)** Cross-polarimetric signature of fine mode.

**Fig. 25.** Polarimetric signature in the vegetation area.

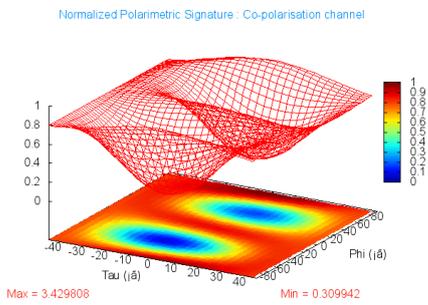

**(a)** Co-polarimetric signature of PSFN.

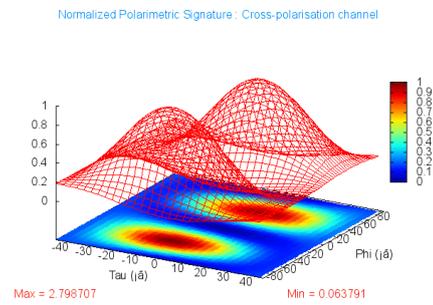

**(b)** Cross-polarimetric signature of PSFN.

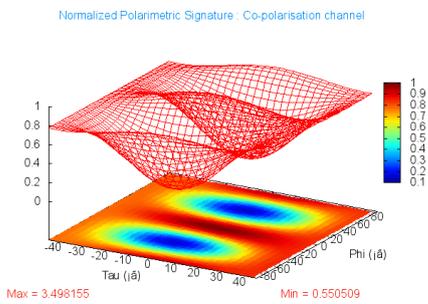

**(c)** Co-polarimetric signature of fine mode.

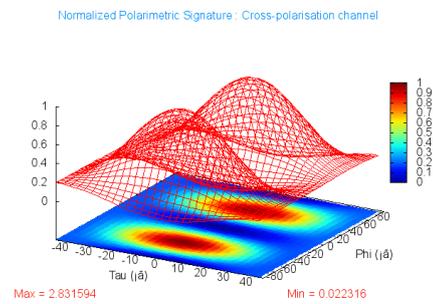

**(d)** Cross-polarimetric signature of fine mode.

**Fig. 26.** Polarimetric signature in the urban built-up area.

Volume scattering is the main scattering form in vegetation regions. For volume scattering, the maximum power of the co-polarimetric signature occurs in the case of linear polarimetric, that is when the ellipticity angle $tau$ is 0. And its maximum power is not related to the polarimetric azimuth. For cross-polarimetric signature, the maximum power appears in the case of circular polarimetric ($tau = \pm 45°$), and the minimum power appears in the case of linear polarimetric. In Fig. 25, the polarimetric signature results of PSFN and HR-PolSAR images in fine mode are consistent, and they are in the volume scattering mode. For urban built-up regions, double scattering accounts for the main part. The minimum power value of the co-polarimetric signature appears in the linear polarimetric case ($tau = 0°$) and the polarimetric azimuth $phi$ is $\pm 45°$. The cross-polarimetric signature graph shows the maximum power value at the corresponding position. As shown in Fig. 26, the polarimetric signature modes of PSFN and HR-PolSAR image in the fine mode in urban built-up areas all conform to the double scattering mechanism.

Among the objects with different polarimetric scattering characteristics, the polarimetric signatures generated by the proposed method conform to the corresponding scattering mechanism. Besides, the polarimetric signature of the proposed fusion network is close to the polarimetric signature result of the corresponding PolSAR image in fine mode.

## 5. CONCLUSION

We propose a PolSAR image and SinSAR image fusion network for generating HR-PolSAR images. Different from the previous PolSAR image super-resolution convolution neural network, we introduced HR-SinSAR images as auxiliary information to the fusion framework, which effectively improved the spatial quality of LR-PolSAR images. To further enhance the feature extraction capability of joint input data, we designed a cross-attention mechanism to extract and cross-weight the dominant features of two input data so that the features of the input data can be fully extracted. Besides, the polarimetric loss function based on the physical imaging mechanism is applied to neural network training, so that the proposed fusion network enhances the spatial texture details and achieves better color fidelity. The experiments on the RADARSAT-2 and AIRSAR datasets prove the effectiveness of PSFN. Compared with the existing PolSAR image resolution enhancement method, the proposed fusion network can effectively use precise spatial information of SinSAR, and it can improve the PolSAR image quality significantly. Besides, the fusion network is easier to converge during training. Due to the limitation of data availability, the weakly supervised or unsupervised deep learning methods for adaptive learning of arbitrary sensors will be the future research trend. Besides, the existing neural networks are black-box models, and the interpretability of the model is poor. Introducing the physical mechanism of PolSAR images into the network is the key to improving network interpretability in the future.